\documentclass[12pt,draftcls,journal,onecolumn]{IEEEtran}
\usepackage{amssymb}
\usepackage{amsfonts}
\usepackage{mathrsfs}
\usepackage{graphicx}
\usepackage{amsmath}
\usepackage{subfigure}
\usepackage{bm}
\usepackage{threeparttable}
\usepackage{dcolumn}
\usepackage{multirow}
\usepackage{cite}
\usepackage{hyperref}
\usepackage[linesnumbered,boxed,vlined]{algorithm2e}
\usepackage{morefloats}
\usepackage{caption}
\usepackage{algorithmicx}
\usepackage{tikz}
\usetikzlibrary{shapes,arrows}
\usepackage{verbatim}

\usepackage{setspace}
\begin{document}

\bibliographystyle{IEEEtran} 

\title{Polar Decomposition of Mutual Information over Complex-Valued Channels}

\author{Qiuliang Xie, Zhaocheng Wang,~\IEEEmembership{Senior
Member,~IEEE}, and Zhixing Yang, ~\IEEEmembership{Senior
Member,~IEEE}
\thanks{Manuscript received July xx, 2012; revised xx xx, 2012. This work was supported by National High Technology Research and Development Program of China (Grant No. 2012AA011704).}
\thanks{Qiuliang Xie  and Zhixing Yang are with the Chinese National Engineering Lab. for Digital TV at Beijing (DTVNEL, Beijing), Beijing 100191, P.R. China (e-mail: xieql@denel.org or xieqiuliang@gmail.com, yangzhx@tsinghua.edu.cn).

Zhaocheng Wang is with the Tsinghua National Laboratory
for Information Science and Technology (TNList), Department of
Electronic Engineering, Tsinghua University, Beijing 100084, P. R.
China (e-mail: zcwang@tsinghua.edu.cn). }}

\maketitle
\begin{abstract}
A polar decomposition of mutual information between a complex-valued channel's input and output is proposed for a input whose amplitude and phase are independent of each other. The mutual information is symmetrically decomposed into three terms: an amplitude term, a phase term, and a cross term, whereby the cross term is negligible at high signal-to-noise ratio. Theoretical bounds of the amplitude and phase terms are derived for additive white Gaussian noise channels with Gaussian inputs. This decomposition is then applied to the recently proposed amplitude phase shift keying with product constellation (product-APSK) inputs. It shows from an information theoretical perspective that coded modulation schemes using product-APSK are able to outperform those using conventional quadrature amplitude modulation (QAM), meanwhile maintain a low complexity.
\end{abstract}


\IEEEpeerreviewmaketitle

\section{Introduction}\label{sec:intro}

For a complex-valued channel, the channel input and output are usually complex-valued signals. Traditionally, the input signal $X$ is decomposed into its real and imaginary parts
\begin{equation*}
X = X_I+j X_Q,
\end{equation*}
where $j=\sqrt{-1}$, and $X_I$ and $X_Q$ denote the real and imaginary parts, also known as the in-phase ($I$) and quadrature ($Q$) parts, respectively. The output signal $Y$ is decomposed as
\begin{equation*}
Y = Y_I+j Y_Q,
\end{equation*}
where $Y_I$ and $Y_Q$ denote the real and imaginary parts of $Y$. Thereafter, the mutual information between $X$ and $Y$ could be decomposed as
\begin{equation}\label{equ:decomposeIQ}
I(X;Y)=I(X_I;Y_I)+I(X_Q;Y_Q|X_I,Y_I)+I(X_I;Y_Q|Y_I)+I(X_Q;Y_I|X_I)
\end{equation}
based on the chain rule of mutual information~\cite[Theorem 2.5.2, Page 24]{xqlRef:Cov06}.

Such decomposition (\ref{equ:decomposeIQ}) can be simplified as
\begin{equation}\label{equ:decomposeIQ_ind}
I(X;Y)=I(X_I;Y_I)+I(X_Q;Y_Q)
\end{equation}
when the following two conditions are satisfied.
\begin{enumerate}
\item{$X_I$ and $X_Q$ are independent of each other, and }
\item{the distortions introduced by the channel affect the real and imaginary parts independently.}
\end{enumerate}

For example, for a rectangular quadrature amplitude modulation (QAM) input over additive white Gaussian noise (AWGN) channels, the channel can be decomposed into two sub-channels, namely, the real and imaginary sub-channels, or say, the $I$ and $Q$ sub-channels~\cite[Page 278]{xqlRef:Pro01}. However, if either of the above two conditions is invalid, the simplified decomposition (\ref{equ:decomposeIQ_ind}) no longer holds. For instance, when a high-order (higher than 4) phase shift keying (PSK) input signal is used, or the channel distortions are $I$-and-$Q$ dependent, e.g., for systems clipping the amplitude caused by non-linear amplifiers, or systems that introduce phase noises.

Most recently, Goebel \textit{et. al} proposed a decomposition of mutual information based on the polar coordinate system, wherein the general case with an arbitrary input is considered, and the mutual information is decomposed into four terms: an amplitude term, a phase term, and two cross terms (called mixed terms therein)~\cite{xqlRef:Goe11}. Such decomposition is helpful in understanding the characteristic of channels with phase noise, and as an example, partially coherent detection was studied therein for fiber-optic communications.

In this correspondence, we investigate the decomposition for a special kind of input whose amplitude and phase are independent of each other, e.g., Gaussian inputs or product amplitude phase shift keying (product-APSK) inputs~\cite{xqlRef:Xie12TWC}. Different from~\cite{xqlRef:Goe11}, with the property of independent amplitude and phase, we symmetrically decompose the mutual information into three terms: an amplitude term, a phase term, and a cross term. Rather than the approximations in~\cite{xqlRef:Goe11}, we derive theoretical bounds of the decomposed terms over AWGN channels for Gaussian inputs. We apply this decomposition into product-APSK inputs, and establish an information theoretical foundation to design and analyze product-APSK for coded modulation (CM) schemes. We show from an information-theoretical perspective that CM schemes using product-APSK are able to achieve better performance while maintain a low complexity, comparing with CM schemes using square QAM. Please note that the decomposition in~\cite[Equ. (3)]{xqlRef:Goe11} is nonsymmetric, and the phase term therein is still relevant to the amplitude of the input signal.


It is worth emphasizing that conventionally it seems as if square QAM were the best choice for practical systems~\cite{xqlRef:Cai98,xqlRef:Gof03}. Therefore, almost all the current communication systems use square QAM constellations to achieve high spectrum efficiency when the average transmit power is limited, including the long-term evolution (LTE) and its advanced version LTE-A~\cite{xqlRef:LTE09}, terrestrial digital video broadcasting (DVB-T) and its second generation DVB-T2~\cite{xqlRef:DVBT2}, wireless local area network (WLAN) standards and etc.  Nevertheless, our recent experiments show that well-designed product-APSK outperforms square QAM in terms of error performance while maintains a low complexity~\cite{xqlRef:Yang11IEICE,xqlRef:Liu11,xqlRef:Xie12TWC}. These experiments motivate us to seek for the information theoretical background for product-APSK. That is why we propose the polar decomposition of mutual information in this correspondence.

As product-APSK is the motivation of our polar decomposition, we would like to introduce the road map of its development. APSK is an old modulation technique proposed several decades ago~\cite{xqlRef:Cahn60}, where the radii are equally spaced to maximize the minimum Euclidean distance. Afterwards, owing to its low peak-to-average power ratio (PAPR), APSK has been optimized for satellite communications, e.g., the 2nd generation digital video broadcasting over satellite (DVB-S2)~\cite{xqlRef:DVBS2,xqlRef:Gau06}. However, these APSK constellations are optimized for transmissions that is peak-power limited, while most communication systems are average-power limited. In addition, such optimization is target for independent demapping, i.e. without any feedback from the decoder to the demapper such as in traditional bit-interleaved coded modulation (BICM) schemes~\cite{xqlRef:Zeh92}, whereby QAM with Gray labeling (Gray-QAM) is much better than APSK when average-power rather than peak-power is limited. Furthermore, since the APSK labeling lacks a nice structure, the complexity of its demapper is higher than that of the Gray-QAM demapper.

Nevertheless, inspired by~\cite{xqlRef:Sun93,xqlRef:Som00,xqlRef:Fra01} that in comparison with conventional QAM signals, shaping can be achieved using a constellation with non-uniformly spaced signal points, we showed that well-designed APSK is capable of providing a considerable shaping gain over complex-valued AWGN channels~\cite{xqlRef:Yang11IEICE}. A basic explanation why APSK may obtain a shaping gain over QAM is that only the complex-valued Gaussian distribution achieves the complex-valued AWGN channel capacity when the average-power is limited. Complex Gaussian distribution is circularly symmetric, while square QAM is nonsymmetric but fortunately APSK is. Therefore, by properly assigning the non-uniformly spaced APSK points, the channel output using APSK would exhibit more complex-Gaussian like behavior than that using QAM.

Furthermore, well-designed Gray-labeled APSK (Gray-APSK) also outperforms Gray-QAM in both independent and iterative demapping scenarios in the sense of error performance~\cite{xqlRef:Liu11}. Iterative demapping refers to that iterations are taken between the demapper and decoder, e.g., in BICM-ID schemes~\cite{xqlRef:Lix98,xqlRef:Bri98}. The concept of Gray-APSK is extended to product-APSK in~\cite{xqlRef:Xie12TWC}, wherein simplified independent demappers are also derived, ensuring that our product-APSK not only outperforms its QAM counterpart in terms of error performance, but also maintains a low complexity.

As a beneficial application of the proposed polar decomposition of mutual information, this correspondence establishes an information theoretical foundation for product-APSK design and analysis. We shall see why product-APSK achieves better performance while maintains a low complexity from an information-theoretical perspective.

The rest of this correspondence is organized as follows. We propose the polar decomposition of mutual information in  Section~\ref{sec:Polar_decom}. Section~\ref{sec:Gaussian} derives the decomposition for Gaussian inputs, where theoretical bounds are presented. In Section~\ref{sec:productAPSK}, we apply such decomposition to product-APSK inputs, which is beneficial for the product-APSK design as well as the construction of its simplified demapper.  Section~\ref{sec:results} provides the numeric results to verify our analysis for both Gaussian and product-APSK inputs. Finally, conclusions are drawn in Section~\ref{sec:con}.

For the sake of clarity, the following notations are employed throughout this correspondence. Upper-case calligraphic symbols denote sets, e.g., $\mathcal{X}$. Symbols in boldface denote vectors,
e.g., $\mathbf{x}$. Upper-case symbols denote random variables (R.V.s), e.g., $X$, while the corresponding lower-case symbols denote their realizations, e.g., $x$. $P_{X}(x)$ is used for the probability of a discrete event of $X=x$, and $p_X(x)$ is used for the probability density function (PDF) of a continuous R.V. $X$. $P_{Y|X}(y|x)$ represents the conditional probability of $Y=y$ given $X=x$, and $p_{Y|X}(y|x)$ represents the conditional PDF of $Y$ given $X=x$. $\log(\cdot)$ denotes the natural logarithm operation, and $\log_2(\cdot)$ denotes the base 2 logarithm operation. $I(X;Y)$ denotes the mutual information between $X$ and $Y$, and $I(X;Y|Z)$ denotes the conditional mutual information between $X$ and $Y$ given $Z$. $H(X)$ denotes the entropy of a discrete R.V. $X$, and $H(Y|X)$ denotes the conditional entropy of $Y$ given $X$. $h(X)$ denotes the differential entropy of a continuous R.V. $X$, and $h(Y|X)$ denotes the conditional differential entropy of $Y$ given $X$. $\mathbb{E}[\cdot]$ denotes the expectation operation, and $\mathbb{E}_{x}[\cdot]$ denotes the expectation with respect to $x$.

\section{Polar Decomposition of Mutual Information}{\label{sec:Polar_decom}}

Consider a channel with complex-valued input $X$ and output $Y$, which could be expressed in a polar-coordinate system that
\begin{equation}\label{equ:X}
X = X_{||}\cdot \exp(jX_{\angle}), \quad X_{||}\in [0,+\infty), X_{\angle}\in [-\pi,\pi),
\end{equation}
and
\begin{equation}\label{equ:Y}
Y = Y_{||}\cdot \exp(jY_{\angle}), \quad Y_{||}\in [0,+\infty), Y_{\angle}\in [-\pi,\pi),
\end{equation}
where $X_{||}$ and $Y_{||}$ denote the amplitudes of the $X$ and $Y$, respectively, and $X_{\angle}$ and $Y_{\angle}$ denote their corresponding phases. Based on the chain rule of mutual information~\cite[Theorem 2.5.2, Page 24]{xqlRef:Cov06}, we have
\begin{equation}\label{equ:decompose1}
\begin{split}
I(X;Y) &= I(X_{||},X_{\angle};Y_{||},Y_{\angle})\\
&=I(X_{||};Y) + I(X_{\angle};Y|X_{||}).
\end{split}
\end{equation}

We focus on a special input case whose amplitude and phase are independent of each other, e.g., for standard complex-valued Gaussian inputs~\cite{xqlRef:Pro01}, or product-APSK inputs~\footnote{The amplitude and phase of a product-APSK input are independent of each other, since we could verify that $P_{X_{||},X_{\angle}}(x_{||},x_{\angle})=P_{X_{||}}(x_{||})\cdot P_{X_{\angle}}(x_{\angle})$, see  Section~\ref{sec:productAPSK} for detail.}. When $X_{||}$ is independent of $X_{\angle}$, we have $h(X_{\angle}|X_{||})=h(X_{\angle})$ for a continuous $X_{\angle}$, or $H(X_{\angle}|X_{||})=H(X_{\angle})$ for a discrete $X_{\angle}$. Nonetheless, by assuming $X_{\angle}$ is continuous without loss of generality, we have
\begin{equation}\label{equ:decompose2}
\begin{split}
I(X_{\angle};Y|X_{||})&=h(X_{\angle}|X_{||})-h(X_{\angle}|X_{||},Y)\\
&=h(X_{\angle})-h(X_{\angle}|Y)+h(X_{\angle}|Y)-h(X_{\angle}|X_{||},Y)\\
&=I(X_{\angle};Y)+I(X_{||};X_{\angle}|Y).
\end{split}
\end{equation}
Therefore, by applying (\ref{equ:decompose2}) to (\ref{equ:decompose1}) we get the decomposition that
\begin{equation}\label{equ:IXY_decomposed}
I(X;Y) = {\underbrace{I(X_{||};Y)}_{\textrm{Amplitude term}}} + \underbrace{I(X_{\angle};Y)}_{\textrm{Phase term}} + \underbrace{I(X_{||};X_{\angle}|Y)}_{\textrm{Cross term}}
\end{equation}
when $X_{||}$ and $X_\angle$ are independent of each other. Please note that our decomposition (\ref{equ:IXY_decomposed}) is different from Goebel's \cite[Equ. (3)]{xqlRef:Goe11} that our phase term $I(X_{\angle};Y)$ is independent of the amplitude of the input signal, and thus we have a nice symmetric expression.


The polar mutual information decomposition (\ref{equ:IXY_decomposed}) is helpful in understanding the characteristic of channels with the input whose amplitude and phase are independent of each other. Traditionally, for square QAM inputs we  decompose the channel into two independent  $I$ and $Q$ sub-channels in order to simplify the detection complexity, when the two conditions shown in Section~\ref{sec:intro} are satisfied. However, product-APSK inputs clearly violate condition 1. Fortunately, by using our polar decomposition (\ref{equ:IXY_decomposed}), we can approximately decompose the channel $C:X \mapsto Y$ into two sub-channels, i.e. the amplitude sub-channel $C_{||}:X_{||}\mapsto Y$, and the phase sub-channel $C_{\angle}:X_{\angle}\mapsto Y$, since we will illustrate that the cross term $I(X_{||};X_{\angle}|Y)$ is negligible. This channel decomposition helps us to simplify the product-APSK demapper in CM schemes.

We now apply the decomposition (\ref{equ:IXY_decomposed}) to the complex-valued AWGN channel
\begin{equation}\label{equ:AWGN_channel_model}
Y = X + W,
\end{equation}
where $Y$ denotes the output signal, $X$ denotes the input signal with power constraint that $\mathbb{E}[|X|^2] = E_s$, and $W$ denotes the Gaussian noise with zero mean and variance of $N_0$ that $W\sim \textrm{CN}(0,N_0)$. The signal-to-noise ratio (SNR) is defined as
\begin{equation}\label{equ:SNR}
\textrm{SNR} = E_s/N_0.
\end{equation}

\section{Gaussian Inputs}\label{sec:Gaussian}
For a complex-valued circularly symmetric Gaussian input that $X\sim \textrm{CN}(0,E_s)$, we derive the expression for each term in (\ref{equ:IXY_decomposed}). Our expression of the amplitude term is quite similar to that in~\cite{xqlRef:Goe11}, except that we derive its lower bound while an approximation was presented in~\cite{xqlRef:Goe11}.

\subsection{The Amplitude Term}
We write the amplitude term $I(X_{||};Y)$ as
\begin{equation}\label{equ:IXaY}
\begin{split}
I(X_{||};Y) &= I(X_{||};Y_{||}) + I(X_{||};Y_{\angle}|Y_{||})\\
&\overset{(a)}{=}I(X_{||};Y_{||})
\end{split}
\end{equation}
by using the chain rule of mutual information~\cite[Theorem 2.5.2, Page 24]{xqlRef:Cov06}, wherein $(a)$ follows from the fact that for a complex-valued Gaussian input $X$, the output $Y$ is also complex-valued Gaussian distributed, and therefore the phase $Y_\angle$ is uniformly distributed within $[-\pi,\pi)$ no matter given the amplitude or not, that is $p_{Y_{\angle}|Y_{||}}(y_{\angle}|y_{||}) = p_{Y_{\angle}|X_{||},Y_{||}}(y_{\angle}|x_{||},y_{||})=p_{Y_\angle}(y_{\angle}) = 1/(2\pi)$ for $y_{\angle}\in [-\pi,\pi)$ and 0 outside, so that we have $I(X_{||};Y_{\angle}|Y_{||})= h(Y_{\angle}|Y_{||}) - h(Y_{\angle}|X_{||}, Y_{||})=0$.

As shown in~\cite{xqlRef:Goe11}, $I(X_{||};Y_{||})$ can be expressed as
\begin{equation}\label{equ:IXaYa_definition}
I(X_{||};Y_{||}) = h(Y_{||}) - h(Y_{||}|X_{||}),
\end{equation}
where~\cite{xqlRef:Goe11}
\begin{equation}\label{equ:hYa}
h(Y_{||})=\frac{1}{2}\log_2(E_s+N_0)+(1+\gamma/2)\log_2 e-1,
\end{equation}
and
\begin{equation}\label{equ:hYaXa}
h(Y_{||}|X_{||}) = -\int_{y_{||}}\int_{x_{||}} p_{X_{||}}(x_{||}) p_{Y_{||}|X_{||}}(y_{||}|x_{||}) \log_2 p_{Y_{||}|X_{||}}(y_{||}|x_{||}) dx_{||} dy_{||}.
\end{equation}
Here in (\ref{equ:hYa}), $\gamma\approx 0.5772$ denotes the \textit{Euler constant}, and the conditional PDF $p_{Y_{||}|X_{||}}(y_{||}|x_{||})$ follows a Rice distribution that~\cite[Page 46]{xqlRef:Pro01}
\begin{equation}\label{equ:pyaxa}
p_{Y_{||}|X_{||}}(y_{||}|x_{||}) = \frac{2y_{||}}{N_0}\cdot \exp\left(-\frac{x_{||}^2+y_{||}^2}{N_0}\right)\cdot I_0\left(\frac{2 x_{||} y_{||}}{N_0}\right),
\end{equation}
where $I_0(\cdot)$ denotes the modified Bessel function of the first kind with order zero.

Clearly (\ref{equ:hYaXa}) does not have a closed form expression, and an approximation was derived in~\cite{xqlRef:Goe11}. In this correspondence, we determine its lower bound. We commence by determining the bound of the conditional variance of $Y_{||}$ given $X_{||}$. We have the first moment of $Y_{||}$ given $X_{||}=x_{||}$ as
\begin{equation}
\begin{split}
\mathbb{E}[Y_{||}|X_{||}=x_{||}] &= \int_0^{\infty}y_{||}p_{Y_{||}|X_{||}}(y_{||}|x_{||}) dy_{||}\\
&=\frac{\sqrt{\pi}}{2\sqrt{N_0}} \exp\left(-\frac{x_{||}^2}{2N_0}\right)\left[(x_{||}^2+N_0)I_0\left(\frac{x_{||}^2}{2N_0}\right)
+x_{||}^2 I_1\left(\frac{x_{||}^2}{2N_0}\right)\right],
\end{split}
\end{equation}
where $I_1(\cdot)$ represents the modified Bessel function of the first kind with order one. We have the second moment of $Y_{||}$ given $X_{||}=x_{||}$ as
\begin{equation}
\begin{split}
\mathbb{E}[Y_{||}^2|X_{||}=x_{||}] &= \int_0^{\infty}y_{||}^2p_{Y_{||}|X_{||}}(y_{||}|x_{||}) dy_{||}\\
&=N_0+x_{||}^2.
\end{split}
\end{equation}
Therefore, the variance of $Y_{||}$ given $X_{||}=x_{||}$ can be evaluated as
\begin{equation}
\begin{split}
\textrm{Var}[Y_{||}|X_{||}=x_{||}] &= \mathbb{E}[Y_{||}^2|X_{||}=x_{||}] - (\mathbb{E}[Y_{||}|X_{||}=x_{||}])^2 \\
&=N_0\underbrace{\left(1+\lambda-\frac{\pi}{4} \exp(-\lambda)\left[(1+\lambda)I_0(\lambda/2)
+\lambda I_1(\lambda/2)\right]^2\right)}_{f(\lambda)},
\end{split}
\end{equation}
where $\lambda \triangleq x_{||}^2/N_0$. As shown in Appendix~\ref{app:proof1}, we have $f(\lambda)<1/2$, and accordingly we have
\begin{equation}
\textrm{Var}[Y_{||}|X_{||}=x_{||}] < N_0/2.
\end{equation}
For a R.V. with a limited variance, the Gaussian distribution maximizes the differential entropy~\cite[p.411, Example 12.2.1]{xqlRef:Cov06} that
\begin{equation}
h(Y_{||}|X_{||}=x_{||})<h(N(0,N_0/2))=\frac{1}{2}\log_2(\pi e N_0)
\end{equation}
$\forall x_{||}\in [0,\infty)$. Now, we have
\begin{equation}\label{equ:hYaXa_upperbound}
h(Y_{||}|X_{||})=\mathbb{E}_{x_{||}}h(Y_{||}|X_{||}=x_{||})<\frac{1}{2}\log_2(\pi e N_0).
\end{equation}
Consequently by applying (\ref{equ:hYaXa_upperbound}), (\ref{equ:hYa}), and (\ref{equ:IXaYa_definition}) into (\ref{equ:IXaY}), we have the lower bound of $I(X_{||};Y)$ that
\begin{equation}\label{equ:lowerbound}
I(X_{||};Y)=I(X_{||};Y_{||})>\frac{1}{2}\log_2(1+\frac{E_s}{N_0}) + \underbrace{\frac{1+\gamma}{2}\log_2 e - \frac{\log_2\pi}{2} - 1}_{\approx -0.69}.
\end{equation}

For a very high SNR, since we have $\lambda=x_{||}^2/N_0\rightarrow \infty$, and $I_0(\lambda/2)\approx \exp(\lambda/2)/\sqrt{\pi\lambda}$~\cite[p.377 9.7.1]{xqlRef:Abr72}, it shows that (\ref{equ:lowerbound}) is also a good approximation~\cite{xqlRef:Goe11}, in other words, the lower bound (\ref{equ:lowerbound}) is tight at high SNR. In fact, we can also show that as $\lambda \rightarrow \infty$, we have $f(\lambda)\rightarrow 1/2$, so that the variance of $Y_{||}$ given $X_{||}$ approaches $N_0/2$, see Appendix~\ref{app:proof1} for the proof.

\subsection{The Phase Term}
The phase term $I(X_{\angle};Y)$ can be written as
\begin{equation}\label{equ:IXpY}
\begin{split}
I(X_{\angle};Y) &= I(X_{\angle};Y_{||}) + I(X_{\angle};Y_{\angle}|Y_{||})\\
&\overset{(a)}{=}I(X_{\angle};Y_{\angle}|Y_{||}),
\end{split}
\end{equation}
where $(a)$ follows from the fact that the output's amplitude $Y_{||}$ is independent of the input's phase $X_{\angle}$ so that we have $I(X_{\angle};Y_{||})=0$. It is notable that our phase term is different from the one in~\cite{xqlRef:Goe11}, wherein it is conditioned on the amplitude of the input signal. Our conditional mutual information $I(X_{\angle};Y_{\angle}|Y_{||})$ can be evaluated as
\begin{equation}\label{equ:IxpYp_Ya_definition}
I(X_{\angle};Y_{\angle}|Y_{||}) = h(Y_{\angle}|Y_{||}) - h(Y_{\angle}|X_{\angle},Y_{||}).
\end{equation}

As the output signal $Y$ is complex-valued Gaussian distributed that $Y\sim \textrm{CN}(0,E_s+N_0)$, the angle $Y_{\angle}$ is uniformly distributed within $[-\pi,\pi)$ meanwhile being independent of $Y_{||}$. Thereby, we have
\begin{equation}\label{equ:Hypya}
h(Y_{\angle}|Y_{||})=h(Y_{\angle})=\log_2(2\pi).
\end{equation}

Since $h(Y_{\angle}|Y_{||},X_{\angle})$ is unaffected by the constant phase shift $X_{\angle}$, we assume $X_{\angle}=0$ without loss of generality, and accordingly, we have
\begin{equation}
\begin{split}
p_{Y_{\angle}|Y_{||},X_{\angle}}&(y_{\angle}|y_{||},x_{\angle}=0)p_{Y_{||}|X_{\angle}}(y_{||}|x_{\angle}=0) =p_{Y_{||},Y_{\angle}|X_{\angle}}(y_{||},y_{\angle}|x_{\angle}=0)\\
&=\int_{0}^{\infty}p_{X_{||}}(x_{||})p_{Y_{||},Y_{\angle}|X_{||},X_{\angle}} (y_{||},y_{\angle}|x_{||},x_{\angle}=0)dx_{||}\\
&=\int_{0}^{\infty}\frac{2 x_{||}}{E_s}\exp\left(-\frac{x_{||}^2}{E_s}\right)\cdot \frac{y_{||}}{\pi N_0} \exp\left(-\frac{x_{||}^2+y_{||}^2-2x_{||}y_{||}\cos y_{\angle}}{N_0}\right)dx_{||}.
\end{split}
\end{equation}
Moreover, since $Y_{||}$ is independent of $X_{\angle}$, the conditional PDF $p_{Y_{||}|X_{\angle}}(y_{||}|x_{\angle}=0)=p_{Y_{||}}(y_{||})$ also follows a Rayleigh distribution. Therefore, we have
\begin{equation}\label{equ:pyp_yaxp}
\begin{split}
p_{Y_{\angle}|Y_{||},X_{\angle}}&(y_{\angle}|y_{||},x_{\angle}=0)= \frac{1}{2\pi}\exp\left[-\frac{y_{||}^2}{N_0(1+\eta)}\right]\\
&+\frac{y_{||}\cos y_{\angle}}{2\sqrt{\pi N_0(1+\eta)}}\exp\left[-\frac{y_{||}^2\sin^2 y_\angle}{N_0(1+\eta)}\right]\cdot\left[1+\textrm{Erf}\left(\frac{y_{||}\cos y_\angle}{\sqrt{N_0(1+\eta)}}\right)\right],
\end{split}
\end{equation}
where $\eta=N_0/E_s$ denotes the inverse of the SNR, and the error function $\textrm{Erf}(x)$ is defined as
\begin{equation}
\textrm{Erf}(x) = \frac{2}{\sqrt{\pi}}\int_0^x \exp(-t^2)dt.
\end{equation}

For a high SNR we have $y_{||}^2/N_0 \rightarrow \infty$, $\eta\rightarrow 0$, and $y_{\angle}\rightarrow x_{\angle}= 0$, and thereby (\ref{equ:pyp_yaxp}) can be approximated as
\begin{equation}\label{equ:pyp_yaxp_approx}
p_{Y_{\angle}|Y_{||},X_{\angle}}(y_{\angle}|y_{||},x_{\angle}=0)\approx \frac{1}{\sqrt{\pi N_0/y_{||}^2}}\exp\left(-\frac{y_{\angle}^2}{N_0/y_{||}^2}\right).
\end{equation}
Since $p(y_{\angle}|y_{||},x_{\angle}=0)$ tends to be the PDF of a real-valued Gaussian distribution with zero mean and variance of $N_0/(2y_{||}^2)$, we obtain
\begin{equation}
\begin{split}
h(Y_\angle|X_\angle,Y_{||}=y_{||})&=h(Y_\angle|X_\angle=0,Y_{||}=y_{||})\\
&\approx \frac{1}{2}\log_2\left(\pi e \cdot \frac{N_0}{y_{||}^2}\right).
\end{split}
\end{equation}
By taking the expectation with respect to $y_{||}$ we have
\begin{equation}\label{equ:hYpGXpYa}
\begin{split}
h(Y_\angle|X_\angle,Y_{||})&=\int_0^{\infty} p_{Y_{||}}(y_{||}) h(Y_\angle|X_\angle=0,Y_{||}=y_{||}) dy_{||}\\
&\approx \int_0^{\infty} \frac{2y_{||}}{E_s+N_0}\exp\left(-\frac{y_{||}^2}{E_s+N_0}\right)\cdot \frac{1}{2}\log_2\left(\pi e \cdot \frac{N_0}{y_{||}^2}\right)dy_{||}\\
&=\frac{1}{2}\log_2\left(\frac{N_0}{E_s+N_0}\right)+\frac{1+\gamma}{2}\log_2 e + \frac{1}{2}\log_2 \pi.
\end{split}
\end{equation}
Applying (\ref{equ:hYpGXpYa}) and (\ref{equ:Hypya}) to (\ref{equ:IxpYp_Ya_definition}) yields
\begin{equation}\label{equ:IXpYp_Ya}
\begin{split}
I(X_\angle;Y_\angle|Y_{||}) &= h(Y_\angle|Y_{||})-h(Y_\angle|X_\angle,Y_{||})\\
&\approx \frac{1}{2}\log_2\left(1+\frac{E_s}{N_0}\right)-\frac{1+\gamma}{2}\log_2 e + \frac{1}{2}\log_2 \pi + 1.
\end{split}
\end{equation}

Based on the lower bound of $I(X_{||};Y)$ shown in (\ref{equ:lowerbound}), the decomposition (\ref{equ:IXY_decomposed}), and the fact that the channel capacity of an AWGN channel is achieved as $I(X;Y)=\log_2(1+E_s/N_0)$ with a Gaussian input, it is clear that (\ref{equ:IXpYp_Ya}) is an upper bound of $I(X_\angle;Y_\angle|Y_{||})$
\begin{equation}\label{equ:upperbound}
I(X_{\angle};Y)=I(X_\angle;Y_\angle|Y_{||}) < \frac{1}{2}\log_2\left(1+\frac{E_s}{N_0}\right)\underbrace{-\frac{1+\gamma}{2}\log_2 e + \frac{1}{2}\log_2 \pi + 1}_{\approx 0.69}.
\end{equation}
This upper bound is also tight at a high SNR shown in (\ref{equ:IXpYp_Ya}).

\subsection{The Cross Term}
The cross term $I(X_{||};X_{\angle}|Y)=I(X_{||};X_{\angle}|Y_{||},Y_{\angle})$ could be calculated as
\begin{equation}\label{equ:IXaXpGYaYp}
I(X_{||};X_{\angle}|Y_{||},Y_{\angle}) = \mathbb{E}_{x_{||},x_{\angle},y_{||},y_{\angle}} \log_2\left[\frac{p_{X_{||},X_{\angle}|Y_{||},Y_{\angle}}(x_{||},x_{\angle}|y_{||},y_{\angle})} {p_{X_{||}|Y_{||},Y_{\angle}}(x_{||}|y_{||},y_{\angle}) p_{X_{\angle}|Y_{||},Y_{\angle}}(x_{\angle}|y_{||},y_{\angle})}\right].
\end{equation}
We have the PDF $p_{X_{||},X_{\angle},Y_{||},Y_{\angle}}(x_{||},x_{\angle},y_{||},y_{\angle})$ expressed as
\begin{equation}\label{equ:pxaxpyayp}
\begin{split}
p_{X_{||},X_{\angle},Y_{||},Y_{\angle}}&(x_{||},x_{\angle},y_{||},y_{\angle}) = p_{X_{||},X_{\angle}}(x_{||},x_{\angle}) p_{Y_{||},Y_{\angle}|X_{||},X_{\angle}}(y_{||},y_{\angle}|x_{||},x_{\angle})\\
&=\frac{x_{||}}{\pi E_s}\exp\left(-\frac{x_{||}^2}{E_s}\right)\cdot
\frac{y_{||}}{\pi N_0}\exp\left[-\frac{x_{||}^2+y_{||}^2-2 x_{||}y_{||}\cos(y_{\angle}- x_{\angle})}{N_0}\right].
\end{split}
\end{equation}
Then, we have the conditional PDF $p_{X_{||},X_{\angle}|Y_{||},Y_{\angle}}(x_{||},x_{\angle}|y_{||},y_{\angle})$ as
\begin{equation}\label{equ:pxaxpGyayp}
p_{X_{||},X_{\angle}|Y_{||},Y_{\angle}}(x_{||},x_{\angle}|y_{||},y_{\angle}) = \frac{E_s\!+\!N_0}{\pi E_s N_0} x_{||}\exp\left[\frac{y_{||}^2}{E_s+N_0}\! -\! \frac{x_{||}^2}{E_s} \!-\! \frac{x_{||}^2+y_{||}^2-2x_{||}y_{||}\cos(y_{\angle}-x_{\angle})}{N_0}\right].
\end{equation}
Now, the conditional PDFs $p_{X_{||}|Y_{||},Y_{\angle}}(x_{||}|y_{||},y_{\angle})$ and $p_{X_{\angle}|Y_{||},Y_{\angle}}(x_{\angle}|y_{||},y_{\angle})$, respectively, can be obtained as
\begin{equation}\label{equ:pxaGyayp}
\begin{split}
p_{X_{||}|Y_{||},Y_{\angle}}(x_{||}|y_{||},y_{\angle}) &= \int_{-\pi}^{\pi}p_{X_{||},X_{\angle}|Y_{||},Y_{\angle}}(x_{||},x_{\angle}|y_{||},y_{\angle})dx_{\angle}\\
&=\frac{2x_{||} (1+\eta)}{N_0}\exp
\left[-\frac{(1+\eta)^2 x_{||}^2+y_{||}^2}{N_0 (1+\eta)}\right]I_0 \left(\frac{2x_{||}y_{||}}{N_0}\right),
\end{split}
\end{equation}
and
\begin{equation}\label{equ:pxpGyayp}
\begin{split}
p_{X_{\angle}|Y_{||},Y_{\angle}}(x_{\angle}|y_{||},y_{\angle}) &= \int_{0}^{\infty}p_{X_{||},X_{\angle}|Y_{||},Y_{\angle}}(x_{||},x_{\angle}|y_{||},y_{\angle})dx_{||}\\
&=\frac{1}{2\pi}\exp\left(-\frac{y_{||}^2}{N_0(1+\eta)}\right)+ \frac{y_{||}\cos(y_{\angle}-x_{\angle})}{2\sqrt{\pi N_0(1+\eta)}} \\ &\times\exp\left[-\frac{y_{||}^2\sin^2(y_{\angle}-x_{\angle})}{N_0(1+\eta)}\right]\cdot
\left[1+\textrm{Erf}\left(\frac{y_{||}\cos(y_{\angle}-x_{\angle})}{\sqrt{N_0(1+\eta)}}\right)\right],
\end{split}
\end{equation}
where $\eta=N_0/E_s$ denotes the inverse of the SNR. It is interesting that the conditional PDF $p_{X_{\angle}|Y_{||},Y_{\angle}}(\cdot|\cdot,\cdot)$ shown in (\ref{equ:pxpGyayp}) is the same as the conditional PDF $p_{Y_{\angle}|Y_{||},X_{\angle}}(\cdot|\cdot,\cdot)$ shown in (\ref{equ:pyp_yaxp}), because the known angle solely affects the centroid.

By applying (\ref{equ:pxaxpGyayp}), (\ref{equ:pxaGyayp}), and (\ref{equ:pxpGyayp}) to (\ref{equ:IXaXpGYaYp}), the cross mutual information $I(X_{||};X_{\angle}|Y_{||},Y_{\angle})$ can be obtained accordingly. However, we do not have a closed-form expression for $I(X_{||};X_{\angle}|Y_{||},Y_{\angle})$. In the following, we discuss two limiting cases either for a very low or a very high SNR.

For a very low SNR that $N_0\rightarrow \infty$, we have
\begin{eqnarray}
p_{X_{||},X_{\angle}|Y_{||},Y_{\angle}}(x_{||},x_{\angle}|y_{||},y_{\angle}) \!\!\!\!\!\!\!\!&&{\rightarrow} \frac{x_{||}}{\pi E_s}\exp\left(-\frac{x_{||}^2}{E_s}\right)\\
p_{X_{||}|Y_{||},Y_{\angle}}(x_{||}|y_{||},y_{\angle}) \!\!\!\!\!\!\!\!&&{\rightarrow} \frac{2 x_{||}}{ E_s}\exp\left(-\frac{x_{||}^2}{E_s}\right)\\
p_{X_{\angle}|Y_{||},Y_{\angle}}(x_{\angle}|y_{||},y_{\angle}) \!\!\!\!\!\!\!\!&&{\rightarrow} \frac{1}{ 2\pi}.
\end{eqnarray}
Therefore, even given $Y_{||}$ and $Y_{\angle}$, the amplitude $X_{||}$ still tends to be Rayleigh distributed, and the phase $X_{\angle}$ tend to be a uniform distribution, and they both tend to be independent of each other, thus we have $I(X_{||};X_{\angle}|Y_{||},Y_{\angle})\rightarrow 0$ for a very low SNR.


For a very large SNR that $N_0\rightarrow 0$, we have $x_{||}\rightarrow y_{||}$ and $x_{\angle}\rightarrow y_{\angle}$. Thereby, we have
\begin{eqnarray}
p_{X_{||},X_{\angle}|Y_{||},Y_{\angle}}(x_{||},x_{\angle}|y_{||},y_{\angle}) \!\!\!\!\!\!\!\!&&{\rightarrow} \frac{1}{\sqrt{\pi N_0}}\exp\!\left[-\frac{(x_{||}-y_{||})^2}{N_0}\right] \frac{1}{\sqrt{\pi N_0/y_{||}^2}}\exp\!\left[-\frac{(x_{\angle}-y_{\angle})^2}{N_0/y_{||}^2}\right]\!\!,\\
p_{X_{||}|Y_{||},Y_{\angle}}(x_{||}|y_{||},y_{\angle}) \!\!\!\!\!\!\!\!&&{\rightarrow} \frac{1}{\sqrt{\pi N_0}}\exp\left[-\frac{(x_{||}-y_{||})^2}{N_0}\right],\\
p_{X_{\angle}|Y_{||},Y_{\angle}}(x_{\angle}|y_{||},y_{\angle}) \!\!\!\!\!\!\!\!&&{\rightarrow} \frac{1}{\sqrt{\pi N_0/y_{||}^2}}\exp\left[-\frac{(x_{\angle}-y_{\angle})^2}{N_0/y_{||}^2}\right],
\end{eqnarray}
for $N_0\rightarrow 0$. In this case, it is interesting that both $X_{||}$ and $X_{\angle}$ tend to be Gaussian distributed given $Y_{||}=y_{||}$ and $Y_{\angle}=y_{\angle}$, i.e., $X_{||}\sim N(y_{||},N_0/2)$ and $X_{\angle}\sim N(y_{\angle},N_0/(2y_{||}^2))$. Moreover, $X_{||}$ and $X_{\angle}$ also tend to be independent of each other even given $Y_{||}$ and $Y_{\angle}$, thus we have $I(X_{||};X_{\angle}|Y_{||},Y_{\angle})\rightarrow 0$ at a very high SNR.

For these two limiting cases, we can also explain the cross terms physically as follows.
For a very low SNR we have $I(X;Y)\rightarrow 0$, and therefore the cross term $I(X_{||};X_{\angle}|Y)$ also tends to 0 because $I(X_{||};X_{\angle}|Y)<I(X;Y)$ according to the decomposition (\ref{equ:IXY_decomposed}). Alternatively, for a very noisy channel, knowing the output $Y$ provides little information about the input $X$, so that we have $I(X_{||};X_{\angle}|Y)\rightarrow I(X_{||};X_{\angle})=0$.

For a very high SNR, we have $Y \rightarrow X$, and therefore we have $I(X_{||};X_{\angle}|Y)\rightarrow I(X_{||};X_{\angle}|X) = 0$. Furthermore, the lower bound of $I(X_{||};Y)$ in (\ref{equ:lowerbound}) and the upper bound of $I(X_\angle;Y)$ in (\ref{equ:upperbound}) are both tight at a high SNR, and it can be observed that
$I(X_{||};Y)+I(X_\angle;Y)\rightarrow \log_2(1+E_s/N_0)=I(X;Y)$. Therefore, it also  shows in this way that $I(X_{||};X_{\angle}|Y)\rightarrow 0$ at a very high SNR.

\section{Product-APSK Inputs}\label{sec:productAPSK}

\begin{figure}[tp]
\centering
\subfigure[]{
\includegraphics[width=0.46\textwidth]{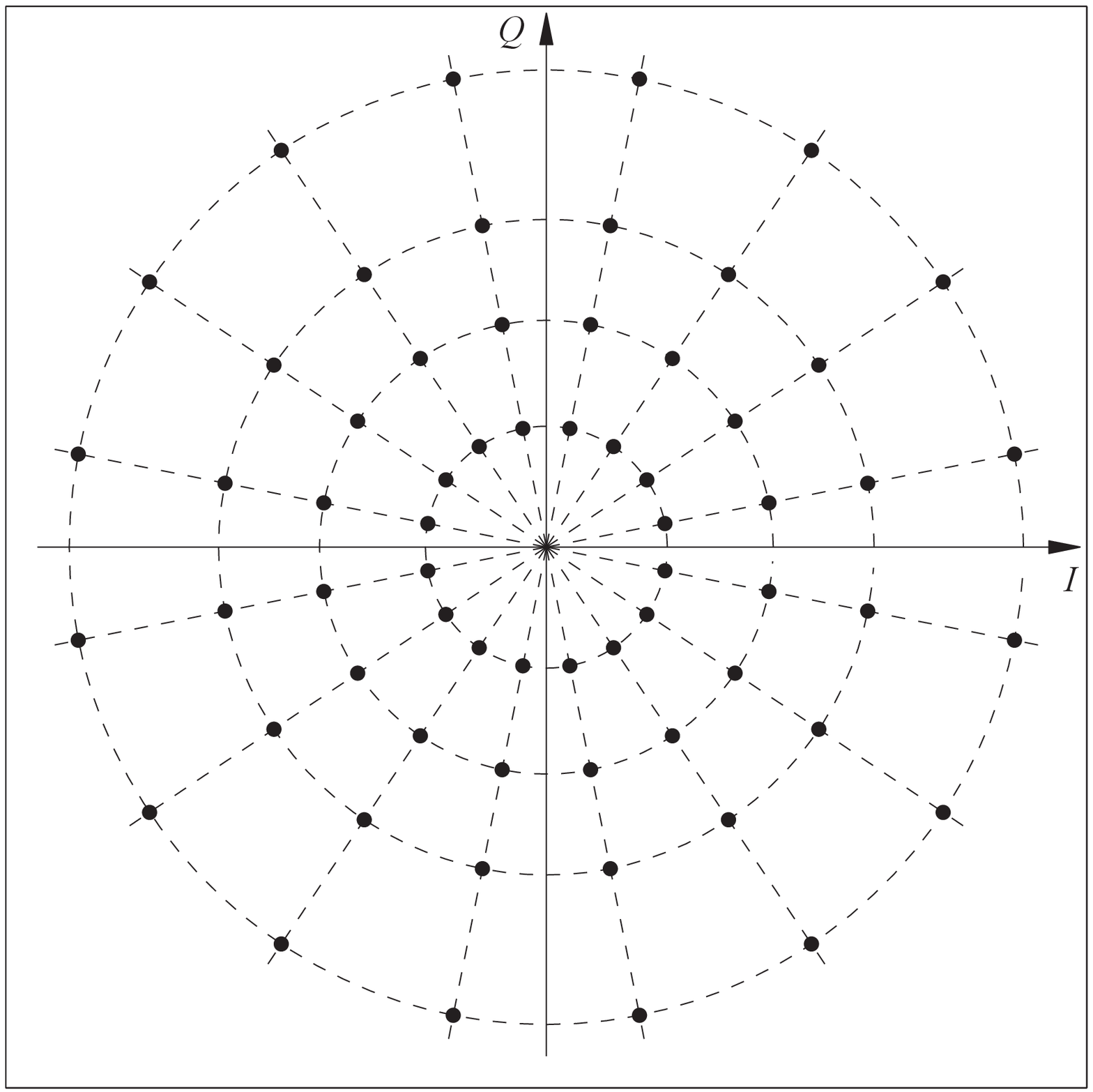}\label{fig:64APSK}
}
\subfigure[]{
\includegraphics[width=0.46\textwidth]{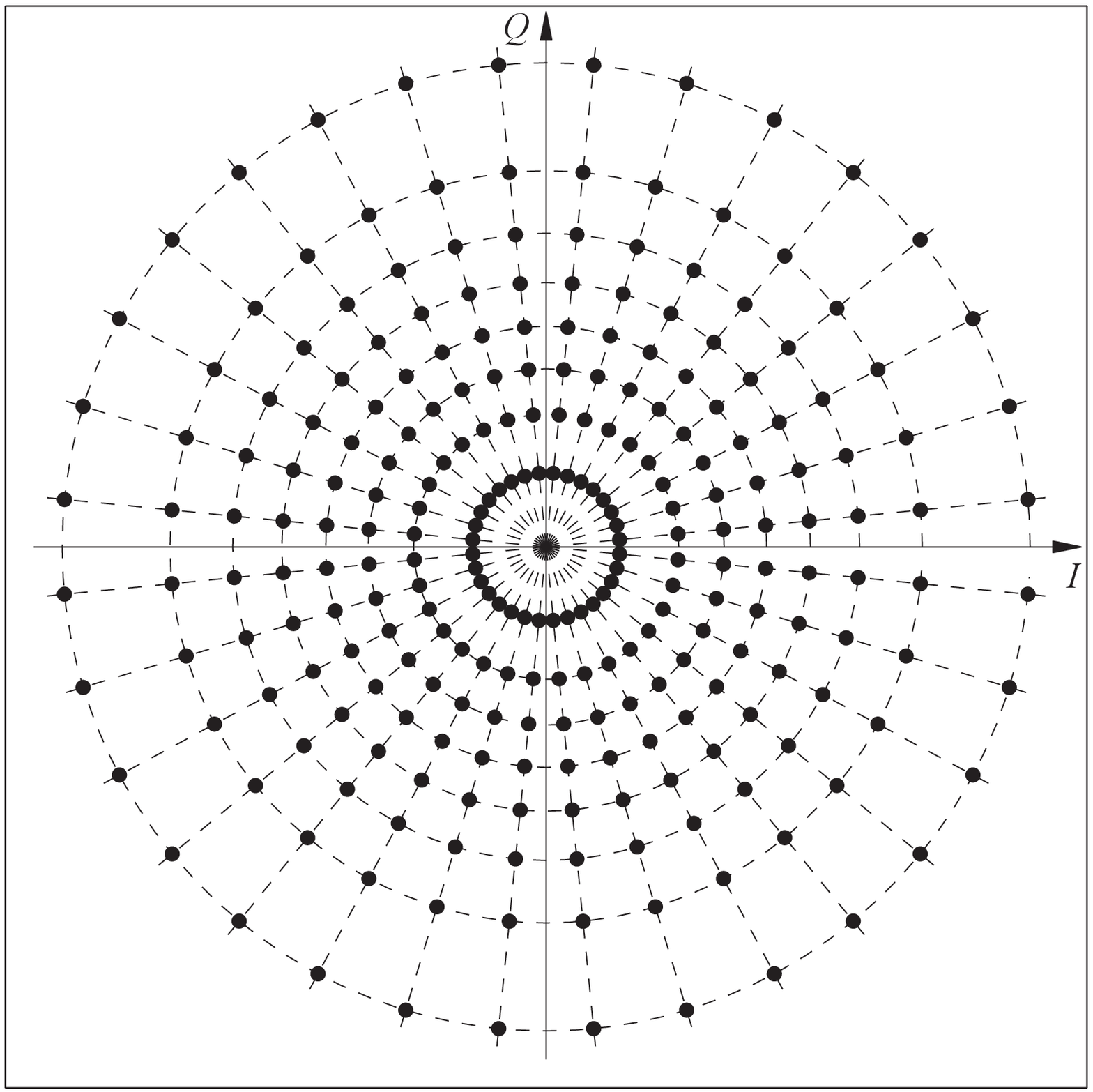}\label{fig:256APSK}
}
\caption{Product-APSK constellations, where the radii are determined according to (\ref{equ:radius}).\\ (a) Product-64APSK, and (b) Product-256APSK.}
\label{fig:APSK}
\end{figure}

Having discussed the decomposition for Gaussian inputs, let us investigate the product-APSK inputs. An $(M=2^{m_{\angle}}\times 2^{m_{||}})$-ary product APSK constellation consists of $2^{m_{||}}$ rings, wherein each ring possesses $2^{m_{\angle}}$ uniformly distributed points. The product-APSK constellation signal set $\mathcal{X}$ is described by
\begin{equation}
\mathcal{X} = \big\{r_{q}\exp(j\varphi_p): p\in\{0,\cdots, 2^{m_{\angle}}-1 \}; q\in\{0,\cdots, 2^{m_{||}}-1\}\big\},
\end{equation}
where $\varphi_p=\frac{\pi}{2^{m_{\angle}}}(2p+1)$ denotes the $p$-th phase-shift, and radius of the $q$-th ring $r_q$ is recommended to be~\cite{xqlRef:Xie12TWC}
\begin{equation}\label{equ:radius}
r_q=\sqrt{-\log\left[1-(q+1/2)\cdot 2^{-m_{||}}\right]}.
\end{equation}
This radius $r_q$ is determined by letting the probability that a standard complex-valued Gaussian R.V. is within the $q$-th ring equal to the probability that the product-APSK signal is within the $q$-th ring, where half the points on the $q$-th ring of the product-APSK are taken into account as within the $q$-th ring. Such radius is quite similar to that for nonuniform PAM design~\cite{xqlRef:Sun93}, or ring constellation design~\cite[Equ. (83)]{xqlRef:Ess10}, whereby the \textit{ring constellation} consists of several rings each with a uniform phase within $[-\pi,\pi)$.

For the parameter pair $(m_{||},m_{\angle})$, we have~\cite{xqlRef:Liu11}
\begin{equation}\label{equ:m}
\left\{
\begin{array}{ll}
m_{\angle} = m/2 + 1, m_{||} = m/2 -1, &\textrm{ for an even } m,\\
m_{\angle} = (m+1)/2, m_{||} = (m-1)/2, &\textrm{ for an odd } m.
\end{array}
\right.
\end{equation}
In~\cite{xqlRef:Liu11}, we determined such pair by maximizing the Harmonic mean of the Euclidean distance. Nonetheless, we could also interpret this assignment according to the decomposition of mutual information in this correspondence. As shown in (\ref{equ:lowerbound}) and (\ref{equ:upperbound}), we have $I(X_{\angle};Y)\approx I(X_{||})+1.38$ (in bits/channel use) for Gaussian inputs at a high SNR. To make the product-APSK more like Gaussian behavior, we shall have $m_{\angle}> m_{||}$ and the gap between them is around 1.38. However, since $m_\angle$ and $m_{||}$ are both integers, the choice (\ref{equ:m}) is reasonable.

It is interesting to note the $(2^{m_{\angle}}\times 2^{m_{||}})$-APSK constellation signal set $\mathcal{X}$ can be regarded as the \textit{product} of the $2^{m_{\angle}}$-PSK set $\mathcal{P}=\{\exp(j\varphi_p)\}$ and the pseudo $2^{m_{||}}$-PAM set $\mathcal{A}=\{r_q\}$, i.e. $\mathcal{X}=\mathcal{P}\times \mathcal{A}$. Additionally, we define the set of the phases as $\mathcal{P}^{\angle}=\{\varphi_p\}$.

Furthermore, based on the product-APSK set, we have the product labeling function $\mu: \mathbf{b}\mapsto x\in\mathcal{X}$, where $\mathbf{b}$ denotes an $m$-bit vector. The function $\mu$ consists of an amplitude-related labeling $\mu_{||}: \mathbf{b}_{||}\mapsto x_{||}\in\mathcal{A}$, and a phase-related labeling $\mu_{\angle}: \mathbf{b}_{\angle}\mapsto  x_{\angle}\in\mathcal{P}^{\angle}$, where $\mathbf{b}_{||}$ denotes an $m_{||}$-bit sub-vector of $\mathbf{b}$ and $\mathbf{b}_{\angle}$ denotes the rest $m_{\angle}$-bit sub-vector, and we have $x=x_{||}\exp(j x_{\angle})$.

According to the above product-APSK constellation labeling, some bits are only relevant to the amplitude of the input signal, and others are only relevant to the phase. Moreover, our recent experiments show that the demapper's complexity could be reduced from the order of $\mathcal{O}(2^m)$ to $\mathcal{O}(2^{m_{||}}+2^{m_{\angle}})$ with a negligible performance loss~\cite{xqlRef:Xie12TWC}. From an information-theoretical perspective, this is because the channel is able to be decomposed into an amplitude sub-channel and a phase sub-channel with a negligible information loss, as detailed below.

We first calculate the mutual information between the product-APSK input and its corresponding output over AWGN channels. The mutual information $I(X;Y)$, with the input $X$ taking on an $M$-ary constellation $\mathcal{X}$ with equal probability, can be evaluated as~\cite{xqlRef:Big05}
\begin{equation}
\begin{split}
I(X;Y) &= \log_2 M - \mathbb{E}_{x,y} \log_2\left[\frac{\sum_{\hat{x}\in\mathcal{X}} p_{Y|X}(y|\hat{x})} {p_{Y|X}(y|x)}\right]\\
&=\log_2 M - \frac{1}{M}\sum_{x\in\mathcal{X}} \mathbb{E}_{w}\log_2 \left[\sum_{\hat{x}\in\mathcal{X}}\exp\left(-\frac{|x-\hat{x}+w|^2-|w|^2}{N_0}\right)\right],
\end{split}
\end{equation}
where $w$ denotes the realization of the complex-valued Gaussian noise $W$ with zero mean and variance of $N_0$. The average symbol energy $E_s$ of the input signal $X$ constrained by the product-APSK constellation $\mathcal{X}$ is determined as
\begin{equation}
E_s = \frac{1}{{2^{m_{||}}}}\sum_{q=0}^{2^{m_{||}}-1} r_q^2.
\end{equation}

It is clear that
\begin{equation}
P_{X_{||},X_{\angle}}(x_{||},x_{\angle})=P_{X_{||}}(x_{||})\cdot P_{X_{\angle}}(x_{\angle})=\frac{1}{M}\delta_{\mathcal{A},\mathcal{P}^{\angle}}(x_{||},x_{\angle})
\end{equation}
for product-APSK inputs, where $\delta_{\mathcal{A},\mathcal{P}^{\angle}}(x_{||},x_{\angle})=1$ if $x_{||}\in\mathcal{A},x_{\angle}\in\mathcal{P}^{\angle}$, and $0$ otherwise. Therefore, the amplitude and phase are independent of each other for product-APSK input, and we also have the polar decomposition (\ref{equ:IXY_decomposed}).

\subsection{The Amplitude Term}

For an $(M=2^{m_{\angle}}\times 2^{m_{||}})$-ary product-APSK input $X$, the amplitude term $I(X_{||};Y)$ can be evaluated as
\begin{equation}
\begin{split}
I(X_{||};Y) &= \mathbb{E}_{x_{||},y}\log_2\left[\frac{p_{Y|X_{||}}(y|x_{||})}{p_{Y}(y)}\right]
=m_{||}+\mathbb{E}_{x_{||},y}\log_2\left[\frac{\displaystyle \sum_{\hat{x}\in\mathcal{X},\hat{x}_{||}=x_{||}} p_{Y|X}(y|\hat{x})}{\displaystyle\sum_{\hat{x}\in\mathcal{X}}p_{Y|X}(y|\hat{x})}\right]\\
&=m_{||} - \frac{1}{M}\sum_{x\in\mathcal{X}} \mathbb{E}_{w}\log_2\left[ \frac{\displaystyle\sum_{\hat{x}\in\mathcal{X}}\exp(-|x-\hat{x}+w|^2/N_0)} {\displaystyle\sum_{\hat{x}\in\mathcal{X},\hat{x}_{||}=x_{||}}\exp(-|x-\hat{x}+w|^2/N_0)} \right].\\
\end{split}
\end{equation}

Furthermore, using the chain rule of mutual information, the amplitude term $I(X_{||};Y)$ can be written as
\begin{equation}
I(X_{||};Y) = I(X_{||};Y_{||}) + I(X_{||};Y_{\angle}|Y_{||}).
\end{equation}
However, unlike the Gaussian input where we have $I(X_{||};Y_{\angle}|Y_{||})=0$ shown in (\ref{equ:IXaY}), the term $I(X_{||};Y_{\angle}|Y_{||})$ usually does not equal to 0 for product-APSK inputs, see Appendix~\ref{app:IxaypGya}. Nevertheless, at a very high SNR that $N_0\rightarrow 0$, we have $Y\rightarrow X$ so that $Y_{||}\rightarrow X_{||}$, and accordingly we have $I(X_{||};Y_{\angle}|Y_{||})\rightarrow 0$, and
\begin{equation}
I(X_{||};Y) \approx I(X_{||};Y_{||}) \approx H(X_{||})=m_{||}.
\end{equation}

In addition, when we have plenty of points on each ring of the product-APSK constellation, we would also have $I(X_{||};Y_{\angle}|Y_{||})\rightarrow 0$, see Appendix~\ref{app:IxaypGya} for the proof. Therefore, we have the approximation $I(X_{||};Y_{||})\approx I(X_{||};Y)$, which suggests that when demapping the bits that are only relevant to the amplitude of the transmitted signal $X$, we can neglect the phase of the received signal $Y$ while only use $Y_{||}$ instead~\cite{xqlRef:Xie12TWC}.

\subsection{The Phase Term}

Similarly, the phase term $I(X_{\angle};Y)$ can be calculated as
\begin{equation}
\begin{split}
I(X_{\angle};Y) &= \mathbb{E}_{x_{\angle},y}\log_2\left[\frac{p_{Y|X_{\angle}}(y|x_{\angle})}{p_{Y}(y)}\right]
=m_{\angle}+\mathbb{E}_{x_{\angle},y}\log_2\left[\frac{\displaystyle \sum_{\hat{x}\in\mathcal{X},\hat{x}_{\angle}=x_{\angle}} p_{Y|X}(y|\hat{x})}{\displaystyle\sum_{\hat{x}\in\mathcal{X}}p_{Y|X}(y|\hat{x})}\right]\\
&=m_{\angle} - \frac{1}{M}\sum_{x\in\mathcal{X}} \mathbb{E}_{w}\log_2\left[ \frac{\displaystyle\sum_{\hat{x}\in\mathcal{X}}\exp(-|x-\hat{x}+w|^2/N_0)} {\displaystyle\sum_{\hat{x}\in\mathcal{X},\hat{x}_{\angle}=x_{\angle}}\exp(-|x-\hat{x}+w|^2/N_0)} \right].
\end{split}
\end{equation}

By using the chain rule of mutual information again, $I(X_{\angle};Y)$ can be written as
\begin{equation}
\begin{split}
I(X_{\angle};Y) &= I(X_{\angle};Y_{||}) + I(X_{\angle};Y_{\angle}|Y_{||})\\
&\overset{(a)}{=}I(X_{\angle};Y_{\angle}|Y_{||}),
\end{split}
\end{equation}
wherein similar to the case of Gaussian inputs, $(a)$ follows from the fact that the output amplitude $Y_{||}$ is independent of the input phase $X_{\angle}$. By the way, at a very high SNR, we have $Y_{\angle}\rightarrow X_{\angle}$ and the following approximation
\begin{equation}
I(X_{\angle};Y) = I(X_{\angle};Y_{\angle}|Y_{||}) \approx H(X_{\angle})=m_{\angle}.
\end{equation}

\subsection{The Cross Term}

The cross term $I(X_{||};X_{\angle}|Y)$ for a product-APSK input can be evaluated as
\begin{equation}
\begin{split}
I(X_{||}&;X_{\angle}|Y)=\mathbb{E}_{y} \sum_{x\in\mathcal{X}} P_{X|Y}(x|y)\log_2 \frac{P_{X|Y}(x|y)}{\displaystyle \sum_{\hat{x}\in \mathcal{X},\hat{x}_{||}=x_{||}}P_{X|Y}(\hat x|y) \sum_{\hat{x}\in \mathcal{X},\hat{x}_{\angle}=x_{\angle}}P_{X|Y}(\hat x|y)}\\
&=\frac{1}{M}\sum_{\tilde{x}\in\mathcal{X}}\sum_{x\in\mathcal{X}}\mathbb{E}_{w} P_{X|Y}(x|\tilde{x}+w)\log_2 \frac{P_{X|Y}(x|\tilde{x}+w)}{\displaystyle\!\!\! \sum_{\hat{x}\in \mathcal{X},\hat{x}_{||}=x_{||}}\!\!\!P_{X|Y}(\hat x|\tilde{x}+w)\!\!\! \sum_{\hat{x}\in \mathcal{X},\hat{x}_{\angle}=x_{\angle}}\!\!\!\!P_{X|Y}(\hat x|\tilde{x}+w)},
\end{split}
\end{equation}
where $P_{X|Y}(x|\tilde{x}+w)$ is expressed as
\begin{equation}
P_{X|Y}(x|\tilde{x}+w) = \frac{\exp(-|\tilde{x}-x+w|^2/N_0)} {\displaystyle\sum_{\hat{x}\in\mathcal{X}}\exp(-|\tilde{x}-\hat{x}+w|^2/N_0)}.
\end{equation}

For a very high SNR, we have $N_0\rightarrow 0, w\rightarrow 0$, and accordingly
\begin{equation}
P_{X|Y}(x|\tilde{x}+w)\approx\left\{
\begin{array}{ll}
1 &\tilde{x}=x\\
0 &\textrm{otherwise}.
\end{array}
\right.
\end{equation}
Therefore, it is clear that $I(X_{||};X_{\angle}|Y)\approx 0$. In addition, since we have $I(X;Y)\rightarrow m_{\angle}+m_{||}$, $I(X_{||};Y)\rightarrow m_{||}$, $I(X_{\angle};Y)\rightarrow m_{\angle}$, and $I(X;Y)=I(X_{||};Y)+I(X_{\angle};Y)+I(X_{||};X_{\angle}|Y)$, we also have the limit that for a very high SNR
\begin{equation}
I(X_{||};X_{\angle}|Y)\rightarrow 0.
\end{equation}

\section{Numeric Results}\label{sec:results}

\subsection{Results of Gaussian Inputs}

\begin{figure}[tp]
\centering
\includegraphics[width=0.7\textwidth]{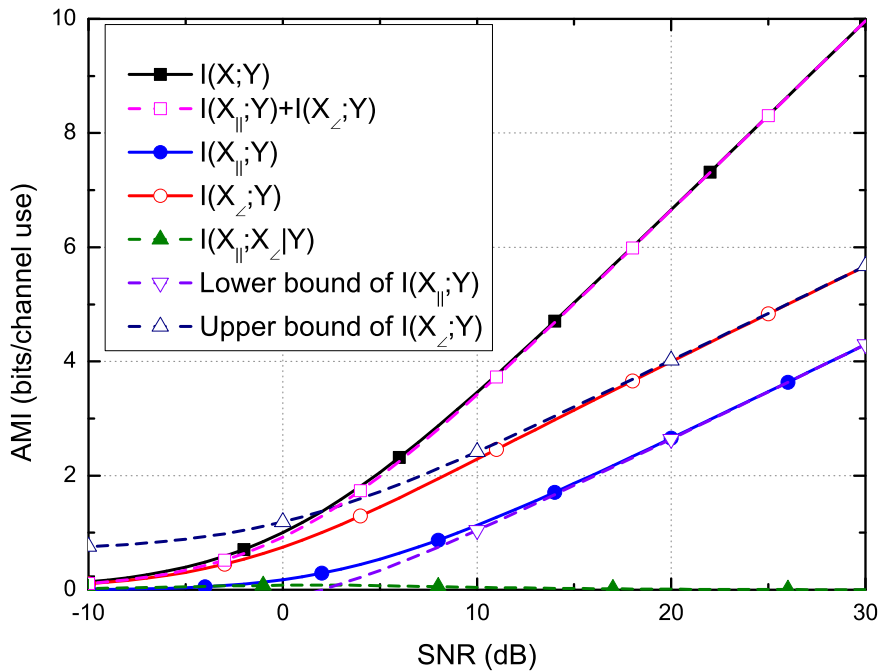}
\caption{Polar decomposed terms of mutual information as a function of SNR for AWGN channels with Gaussian inputs. The lower bound of the amplitude term $I(X_{||};Y)$ shown in (\ref{equ:lowerbound}) and the upper bound of the phase term $I(X_{\angle};Y)$ shown in (\ref{equ:upperbound}) are also depicted. }
\label{fig:GGaussian}
\end{figure}

We now present the results of the decomposed terms of mutual information. The results for AWGN channels with Gaussian inputs are shown in Fig.~\ref{fig:GGaussian}, wherein the notation AMI denotes the average mutual information.  The lower bound of the amplitude term $I(X_{||};Y)$ in (\ref{equ:lowerbound}) and the upper bound of the phase term $I(X_{\angle};Y)$ in (\ref{equ:upperbound}) are also depicted. It shows that these two bounds are both tight for a SNR higher than 12 dB. The cross term $I(X_{||};X_{\angle}|Y)$ reaches its maximum value of about 0.08 bits/channel use at $\textrm{SNR}\approx 1$ dB, and it tends to be zero at a high SNR. Moreover, the cross term is negligible compared to the amplitude or the phase term, which indicates that the AWGN channel can be decomposed into an amplitude sub-channel and a phase sub-channel with a negligible information loss for Gaussian inputs. For example, the gap between $I(X;Y)$ and $I(X_{||};Y)+I(X_{\angle};Y)$, that is, the cross term $I(X_{||};X_{\angle}|Y)$, is less than 0.04 bits/channel use at a SNR over 12 dB, and less than 0.02 bits/channel use at a SNR over 15 dB. In other words, for mutual information higher than 4 bits/channel use, the loss is less than 0.1 dB. For a clearer observation of the cross term, please refer to Fig.~\ref{fig:GdeltaAMI}.


\subsection{Results of Product-APSK Inputs}

\begin{figure}[tp]
\centering
\includegraphics[width=0.7\textwidth]{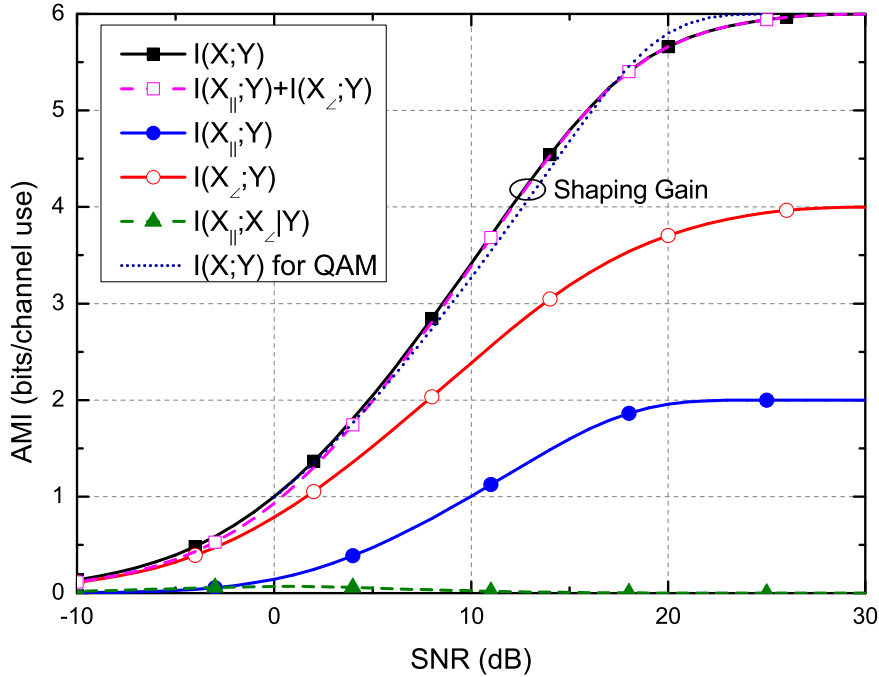}
\caption{Polar decomposed terms of mutual information as a function of SNR for AWGN channels with 64APSK inputs depicted in Fig.~\ref{fig:64APSK}, the AMI associated with 64QAM inputs is also depicted for reference to illustrate the shaping gain obtained by product-APSK.}
\label{fig:AMI64APSK}
\end{figure}


\begin{figure}[tp]
\centering
\includegraphics[width=0.7\textwidth]{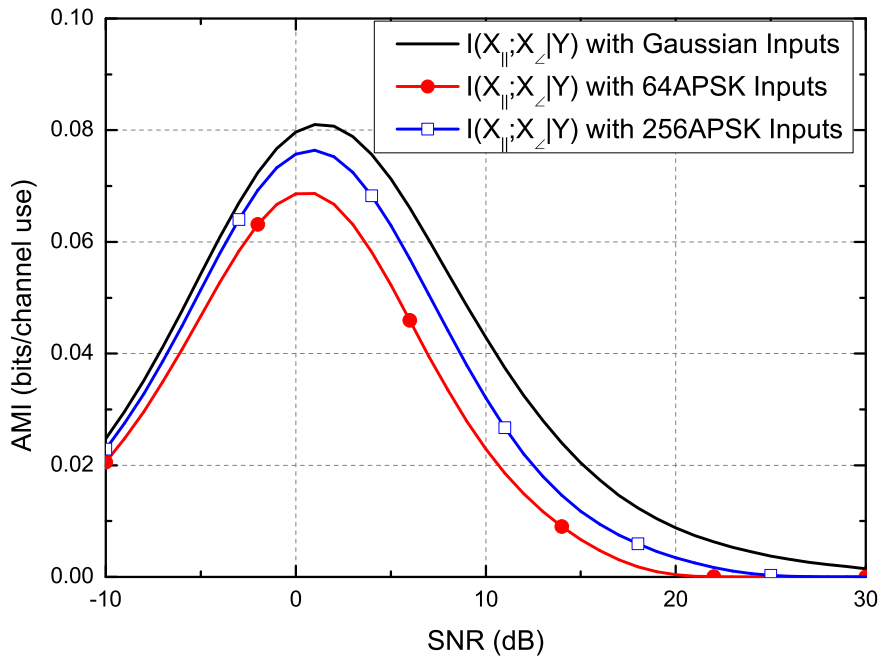}
\caption{The cross term of decompositions as a function of SNR for AWGN channels with Gaussian inputs, and the product-APSK inputs depicted in Fig.~\ref{fig:64APSK} and Fig.~\ref{fig:256APSK}.}
\label{fig:GdeltaAMI}
\end{figure}

We take $(16\times 4=64)$-APSK as an example. The constellation is illustrated in Fig.~\ref{fig:64APSK}, with the radii given by (\ref{equ:radius}). The decomposition results are presented in Fig.~\ref{fig:AMI64APSK}.  These results are quite similar to the Gaussian-input case except that the amplitude term $I(X_{||};Y)$ is upper-bounded by $m_{||}$, and the phase term $I(X_{\angle};Y)$ is upper-bounded by $m_{\angle}$, at high SNRs. The cross term $I(X_{||};X_{\angle}|Y)$ is negligible. For instance, for coding rates higher than 1/2, namely, for the mutual information is higher than 3 bits/channel use for 64APSK inputs, such loss is about 0.1 dB. In addition, the AMI $I(X;Y)$ associated with 64QAM input is also depicted for reference. Fig.~\ref{fig:AMI64APSK} clearly shows that even when decomposition is used, product-64APSK still outperforms 64QAM at code rates of usual interests such as 1/2 or 2/3. For example, about 0.6 dB shaping gain can be obtained at the code rate of 2/3.

We collect the results of the cross term associated with three input cases together in Fig.~\ref{fig:GdeltaAMI}, including Gaussian, 64APSK and 256APSK. It is interesting that all of they reach their maximum value at the $\textrm{SNR}\approx 1$ dB. Moreover, it shows that the cross term increases with the constellation order, and intuitively, the curve associated with Gaussian inputs may be the limit for product-APSK inputs when the constellation order goes to infinity.

Although we only examined two kinds of inputs, namely, the Gaussian input and the product-APSK input. This decomposition is applicable to other inputs with independent amplitude and phase, such as PSK and phase-modulated~\cite[Sec.III(b)]{xqlRef:Goe11} which is named as ring constellation in~\cite{xqlRef:Ess10}. Indeed, PSK can be regarded as a degradation of APSK that consists of a single ring, while the ring constellation is a limiting case of APSK that possesses infinite points on each ring.  Therefore, our decomposition is also applicable to these inputs as a simple extension.

\section{Conclusions}{\label{sec:con}}

We have proposed a novel polar decomposition of mutual information for complex-valued channels with an input whose amplitude and phase are independent of each other. Using this decomposition, the mutual information between the channel's input and output is symmetrically decomposed into three terms: an amplitude term, a phase term, and a cross term, where the cross term is negligible, based on which the channel can be approximately decomposed into two sub-channels associated with amplitude and phase, respectively. This decomposition is then performed for AWGN channels with Gaussian and product-APSK inputs. For Gaussian inputs, theoretical bounds are derived. For product-APSK inputs, the decomposition is helpful to facilitate the design of product-APSK, and directly leads to a simplified demapper. This establishes a solid information theoretical foundation for coded modulation schemes using product-APSK, that is, better performance can be achieved by product-APSK over QAM while the complexity is maintained low.

\begin{appendix}

\subsection{Proof of $ f(\lambda)<1/2$}\label{app:proof1}

We have the definition of $f(\lambda)$ as
\begin{equation}
\begin{split}
f(\lambda)&=1+\lambda - \frac{\pi}{4}\exp(-\lambda)[(1+\lambda)I_0(\lambda/2)+\lambda I_1(\lambda/2)]^2\\
&=1+\lambda - \frac{\pi}{4}[L_{1/2}(-\lambda)]^2,
\end{split}
\end{equation}
where $L_{1/2}(x)=\exp(x/2)[(1-x)I_0(-x/2)-x I_1(-x/2)]$ denotes the Laguerre polynomial with the order of $1/2$. We have~\cite[9.6.10, Page 375]{xqlRef:Abr72}
\begin{equation}
I_v(x) = \sum_{k=0}^{\infty} \frac{(x/2)^{2k+v}}{k!\Gamma(v+k+1)},
\end{equation}
where $\Gamma(\cdot)$ denotes the Gamma function and $\Gamma(n+1)=n!$ for a positive integer $n$.
Therefore, for a positive $x$, we have
\begin{equation}
\begin{split}
I_0(x)= \sum_{k=0}^{\infty} \frac{(x/2)^{2k}}{(k!)^2}> 1 + \frac{x^2}{4}
\end{split}
\end{equation}
and
\begin{equation}
\begin{split}
I_1(x) = \sum_{k=0}^{\infty} \frac{(x/2)^{2k+1}}{k!(k+1)!}> \frac{x}{2}.
\end{split}
\end{equation}
Thereby, we consequently have
\begin{equation}\label{equ:Lbound1}
L_{1/2}(-\lambda)>\exp(-\lambda/2)\left(1+\lambda+\frac{5}{16} \lambda^2\right),
\end{equation}
and
\begin{equation}\label{equ:upperboundf1}
f(\lambda)<1+\lambda -\frac{\pi}{4}\exp(-\lambda)\left(1+\lambda+\frac{5}{16} \lambda^2\right)^2.
\end{equation}
It is easy to verify that the right side of (\ref{equ:upperboundf1}) is an monotone increasing function with $\lambda$. Therefore, for $\lambda\in [0,1]$, we have
\begin{equation}\label{equ:fproofL1}
f(\lambda)\le f(1)= 2 - \frac{1369 \pi}{1024 e} \approx 0.455<1/2, \quad \forall \lambda\in [0,1].
\end{equation}

For $\lambda>1$, we have the series associated with the modified Bessel function of the first kind with order $v$ as~\cite[9.7.1, Page 377]{xqlRef:Abr72}
\begin{equation}
I_v(x)\sim \frac{e^x}{\sqrt{2\pi x}}\left[1-\frac{u-1}{8x}+\frac{(u-1)(u-9)}{2!(8x)^2}-\frac{(u-1)(u-9)(u-25)}{3!(8x)^3}+\cdots\right],
\end{equation}
for a large $x$ where $u=4v^2$. Therefore, we have
\begin{equation}
\begin{split}
I_0(\lambda/2)&\sim \frac{e^{\lambda/2}}{\sqrt{\pi \lambda}} \left[1+\sum_{n=1}^{\infty} \frac{\prod_{k=1}^n (2k-1)^2}{n!4^n}\left(\frac{1}{\lambda}\right)^n \right]\\
&=\frac{e^{\lambda/2}}{\sqrt{\pi \lambda}} \left[1+\sum_{n=1}^{\infty} \frac{\Gamma^2(n+1/2)}{{\pi}n!}\left(\frac{1}{\lambda}\right)^n \right],
\end{split}
\end{equation}
and
\begin{equation}
\begin{split}
I_1(\lambda/2)&\sim \frac{e^{\lambda/2}}{\sqrt{\pi \lambda}} \left[1+\sum_{n=1}^{\infty} \frac{(-1)^n \prod_{k=1}^n [4-(2k-1)^2]}{n!4^n}\left(\frac{1}{\lambda}\right)^n \right]\\
&=\frac{e^{\lambda/2}}{\sqrt{\pi \lambda}} \left[1-\sum_{n=1}^{\infty} \frac{\Gamma(n+3/2)\Gamma(n-1/2)}{{\pi}n!}\left(\frac{1}{\lambda}\right)^n \right].
\end{split}
\end{equation}
Thereby, we have
\begin{equation}\label{equ:Lbound2}
\begin{split}
L_{1/2}(-\lambda)&\sim
\frac{1}{\sqrt{\pi \lambda}} \left[2\lambda +\frac{1}{2}\right.\\
&\left. +\sum_{n=1}^{\infty} \left(\frac{(1+n)\Gamma^2(n+\frac{1}{2})+\Gamma^2(n+\frac{3}{2})-\Gamma(n+\frac{1}{2})\Gamma(n+\frac{5}{2})}{\pi (n+1)!}\right)\left(\frac{1}{\lambda}\right)^n\right]\\
&=\frac{1}{\sqrt{\pi \lambda}} \left[2\lambda +\frac{1}{2}+\sum_{n=1}^{\infty} \left(\frac{\Gamma^2(n+\frac{1}{2})}{2\pi (n+1)!}\right)\left(\frac{1}{\lambda}\right)^n\right]\\
&>\frac{1}{\sqrt{\pi }} \left(2\sqrt{\lambda} +\frac{1}{2\sqrt{\lambda}}\right).
\end{split}
\end{equation}
In the above proof, we have used the Gamma function that
\begin{equation}
\Gamma(n+1/2) = \frac{\prod_{k=1}^{n}(2k-1)}{2^n} \Gamma(1/2)
\end{equation}
and $\Gamma(1/2) = \sqrt{\pi}$.
Now we may write that for $\lambda>1$, we have
\begin{equation}\label{equ:fproofG1}
f(\lambda)<1+\lambda - \frac{\pi}{4}\cdot \frac{1}{\pi}\left(2\sqrt{\lambda}+\frac{1}{2\sqrt{\lambda}}\right)^2<1/2, \quad \forall \lambda\in (1,\infty).
\end{equation}

\begin{figure}[tp]
\centering
\includegraphics[width=0.7\textwidth]{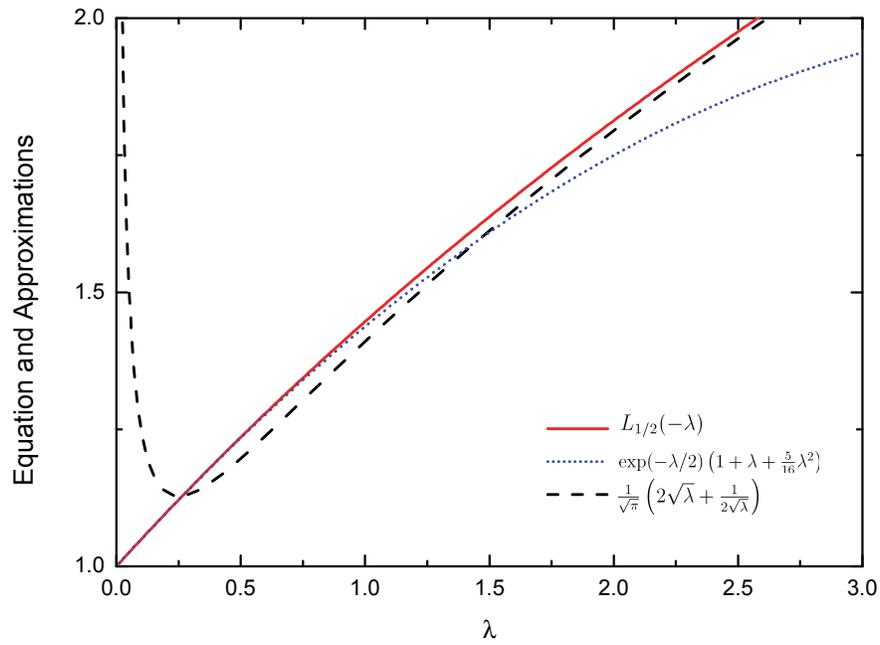}
\caption{The generalized Laguerre polynomial of the order $1/2$ and its approximations.}
\label{fig:LaguerrePoly}
\end{figure}

\begin{figure}[tp]
\centering
\includegraphics[width=0.7\textwidth]{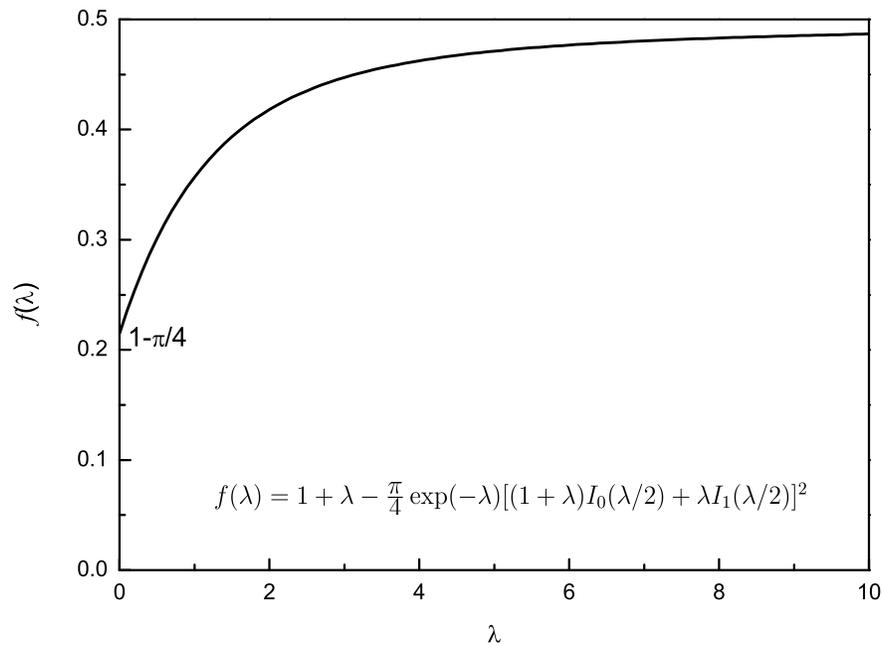}
\caption{Numeric results of $f(\lambda)$ as a function of $\lambda$.}
\label{fig:GRicianVariance}
\end{figure}

Based on (\ref{equ:fproofL1}) and (\ref{equ:fproofG1}), we may have $f(\lambda)<1/2,\forall \lambda\ge 0$. For an intuitive imagination of the above proof, we provide the numeric results of $L_{1/2}(-\lambda)$ and its approximations (\ref{equ:Lbound1}) and (\ref{equ:Lbound2}) in Fig.~\ref{fig:LaguerrePoly}. In addition, $f(\lambda)$ as a function of $\lambda$ is plotted in Fig.~\ref{fig:GRicianVariance}.

\subsection{Proof of $I(X_{||};Y_{\angle}|Y_{||})\neq 0$ for product-APSK inputs }\label{app:IxaypGya}

To show $I(X_{||};Y_{\angle}|Y_{||})\neq 0$, it is equivalent to show that $p_{Y_{\angle}|X_{||},Y_{||}}(y_{\angle}|x_{||},y_{||})\neq p_{Y_{\angle}|Y_{||}}(y_{\angle}|y_{||})$.
It is also equivalent that $p_{Y_{\angle}|X_{||},Y_{||}}(y_{\angle}|x_{||},y_{||})$ is relevant to $x_{||}$. In fact, we have
\begin{equation}
\begin{split}
p_{X_{||},X_{\angle},Y_{||},Y_{\angle}}(x_{||},x_{\angle},y_{||},y_{\angle})&= P_{X_{||},X_{\angle}}(x_{||},x_{\angle}) p_{Y_{||},Y_{\angle}|X_{||},X_{\angle}}(y_{||},y_{\angle}|x_{||},x_{\angle})\\
&=P_{X_{||},X_{\angle}}(x_{||},x_{\angle})\cdot \frac{y_{||}}{\pi N_0} \exp\left[-\frac{x_{||}^2+y_{||}^2-2x_{||}y_{||}\cos(y_{\angle}-x_{\angle})}{N_0}\right].
\end{split}
\end{equation}
The summation of $p_{X_{||},X_{\angle},Y_{||},Y_{\angle}}(x_{||},x_{\angle},y_{||},y_{\angle})$ with respect to $x_{\angle}$ yields
\begin{equation}
p_{X_{||},Y_{||},Y_{\angle}}(x_{||},y_{||},y_{\angle})=\frac{1}{M}\sum_{x_{\angle}} \frac{y_{||}}{\pi N_0} \exp\left[-\frac{x_{||}^2+y_{||}^2-2x_{||}y_{||}\cos(y_{\angle}-x_{\angle})}{N_0}\right]\delta_{\mathcal{A}}(x_{||}),
\end{equation}
where $\delta_{\mathcal{A}}(x_{||})=1$ if $x_{||}\in \mathcal{A}$, and 0 otherwise. $\mathcal{A}$ denotes the set of the amplitudes of our product-APSK defined in Section~\ref{sec:productAPSK}. Furthermore,  by integrating $p_{X_{||},Y_{||},Y_{\angle}}(x_{||},y_{||},y_{\angle})$ with respect to $y_{\angle}$ yields
\begin{equation}
p_{X_{||},Y_{||}}(x_{||},y_{||})=\frac{1}{2^{m_{||}}} \frac{2 y_{||}}{N_0} \exp\left(-\frac{x_{||}^2+y_{||}^2}{N_0}\right)I_0\left(\frac{2x_{||}y_{||}}{N_0}\right)\delta_{\mathcal{A}}(x_{||}).
\end{equation}
Now we have
\begin{equation}\label{equ:pypGxaya_APSK}
\begin{split}
p_{Y_{\angle}|X_{||},Y_{||}}(y_{\angle}|x_{||},y_{||})&= \frac{p_{Y_{\angle},X_{||},Y_{||}}(y_{\angle},x_{||},y_{||})}{p_{X_{||},Y_{||}}(x_{||},y_{||})}\\
&=\frac{1}{2^{m_{\angle}}}\sum_{x_{\angle}}\frac{1}{2\pi} \frac{\exp\left[2x_{||}y_{||}\cos(y_{\angle}-x_{\angle})/N_0\right]}
{I_0(2x_{||}y_{||}/N_0)}\delta_{\mathcal{A}}(x_{||}).
\end{split}
\end{equation}
It is clear that $p_{Y_{\angle}|X_{||},Y_{||}}(y_{\angle}|x_{||},y_{||})$ is relevant to $x_{||}$ based on (\ref{equ:pypGxaya_APSK}), and thereby usually we have $I(X_{||};Y_{\angle}|Y_{||})\neq 0$ for a product-APSK input.

However, at a very high SNR with $N_0\rightarrow 0$, we $y_{\angle}\rightarrow x_{\angle}$, $y_{||}\rightarrow x_{||}$, and by using the approximation that $I_0(z)\approx \exp(z)/\sqrt{2\pi z}$, (\ref{equ:pypGxaya_APSK}) can be approximated as
\begin{equation}
p_{Y_{\angle}|X_{||},Y_{||}}(y_{\angle}|x_{||},y_{||})\approx \frac{1}{2^{m_{\angle}}}\sum_{x_{\angle}}\frac{1}{\sqrt{\pi N_0/x_{||}^2}}\exp\left[-\frac{(y_{\angle}-x_{\angle})^2}{N_0/x_{||}^2}\right] \delta_{\mathcal{A}}(x_{||}),
\end{equation}
That is, the PDF of $Y_{\angle}$ given $X_{||}=x_{||}\in \mathcal{A}$ and $Y_{||}$ is a uniformly weighted combination of a serial Gaussian distribution with the mean of $y_{\angle}\in \mathcal{P}^{\angle}$ (defined in Section~\ref{sec:productAPSK}), and the variance of $N_0/(2x_{||}^2)$. Furthermore, since $y_{||}\rightarrow x_{||}$ at a very high SNR, we also have $p_{Y_{\angle}|X_{||},Y_{||}}(y_{\angle}|x_{||},y_{||})\approx p_{Y_{\angle}|Y_{||}}(y_{\angle}|y_{||})\approx p_{Y_{\angle}|X_{||}}(y_{\angle}|x_{||})$. Therefore, we have $I(X_{||};X_{\angle}|Y)\rightarrow 0$.

In addition, when we have lots of points on each ring that $m_{\angle}\rightarrow \infty$, (\ref{equ:pypGxaya_APSK}) could be simplified as
\begin{equation}\label{equ:pypGxaya_APSK2}
\begin{split}
p_{Y_{\angle}|X_{||},Y_{||}}(y_{\angle}|x_{||},y_{||})
&=\frac{1}{2\pi}\cdot\frac{\delta_{\mathcal{A}}(x_{||})}{2\pi I_0(2x_{||}y_{||}/N_0)}\sum_{x_{\angle}}\frac{2\pi}{2^{m_{\angle}}}\exp[2x_{||}y_{||}\cos(x_{\angle}-y_{\angle})/N_0]\\
&=\frac{1}{2\pi}\cdot\frac{\delta_{\mathcal{A}}(x_{||})}{2\pi I_0(2x_{||}y_{||}/N_0)} \int_{-\pi}^{\pi}\exp[2x_{||}y_{||}\cos(x_{\angle}-y_{\angle})/N_0] dx_{\angle}\\
&=\frac{1}{2\pi}\delta_{\mathcal{A}}(x_{||}).
\end{split}
\end{equation}

Thereby, the phase of the output signal would be uniformly distributed within $[-\pi,\pi)$, meanwhile being independent of its magnitude, for product-APSK inputs when the number of points on each ring tends to infinity (typically, when the constellation order tends to infinity). Intuitively, (\ref{equ:pypGxaya_APSK2}) is reasonable, because when the number of points on each ring tends to infinity, the phase of input signal would also tends to be uniformly distributed within $[-\pi,\pi)$. In this case, we would have $I(X_{||};Y_{\angle}|Y_{||})\rightarrow 0$ which is quite similar to the Gaussian inputs.

\end{appendix}

\bibliography{IEEEabrv,xqlRef}

\end{document}